\documentclass[12pt]{article}
\usepackage[cp1251]{inputenc}
\usepackage[english]{babel}
\usepackage{indentfirst}
\def\a{\alpha}

\def\d{\delta}
\def\e{\epsilon}

\def\p{\psi}

\def\s{\sigma}

\def\G{\Gamma}

\def\be{\begin{equation}}
\def\ee{\end{equation}}
\def\arr{\begin{array}{rll}}
\def\ea{\end{array}}
\def\bea{\begin{eqnarray}}
\def\eea{\end{eqnarray}}

\begin{document}

\begin{titlepage}
\noindent
\renewcommand{\thefootnote}{\fnsymbol{footnote}}
\vskip 1.5cm

\begin{center}

{\Large\bf Making the hyper--K\"ahler structure}\\

\vskip 0.5cm

{\Large\bf  of N=2 quantum string manifest }\\

\bigskip

\vskip 1.5cm
%{\Large
{\large S. Bellucci}${}^a$\footnote{bellucci@lnf.infn.it},
{\large A.V. Galajinsky}${}^b$\footnote{galajin@mph.phtd.tpu.edu.ru} and
{\large E. Latini}${}^a$\footnote{latini@lnf.infn.it}

\vskip 0.2cm

${}^a$ {\it INFN--Laboratori Nazionali di Frascati, C.P. 13,
00044 Frascati, Italy}\\

\vskip 0.2cm
${}^b${\it Laboratory of Mathematical Physics, Tomsk Polytechnic University, \\
634050 Tomsk, Lenin Ave. 30, Russian Federation}

\end{center}

\renewcommand{\thefootnote}{\arabic{footnote}}
\setcounter{footnote}0
\vskip 1.5cm

\begin{abstract}
\noindent
We show that the Lorentz covariant formulation of $N=2$ string in a curved space
reveals an explicit hyper--K\"ahler structure. Apart from the metric, the superconformal
currents couple to a background two--form. By superconformal symmetry the latter is
constrained to be holomorphic and covariantly constant and allows one to construct
three complex structures obeying a (pseudo)quaternion algebra.
\end{abstract}

\vspace{0.5cm}


\end{titlepage}

\noindent

\vskip 0.3cm
{\bf 1. Introduction}

\noindent

\vskip 0.5cm

One of the fascinating features of two--dimensional superconformal symmetry is
that it relates string theory and geometry. Consistent backgrounds where
string theory may propagate are identified with low lying string
states either by analyzing one--loop divergences
of the corresponding string theory effective action \cite{cfmp}--\cite{ft}, or evaluating
operator product expansions of superconformal currents in a curved space \cite{bns},
or studying renormalisation of the trace of the stress--energy tensor \cite{osborn}.

With the number of world-sheet supersymmetries growing, one reveals more refined geometrical structures.
In particular, $N=2$ superconformal symmetry requires a Ricci--flat
K\"ahler space as a consistent background \cite{itoh} which correlates well with the sigma model analysis
of Ref. \cite{agg}. Notice, however, that if one is concerned
with superconformal algebras admitting unitary representations, the restriction $N\leq 4$ on the number of
fermionic currents holds \cite{ramsch}. On the sigma model side the same bound follows from the requirement
that a target manifold is irreducible \cite{agg}.

String theory incorporating $N=2$ superconformal symmetry has the central charge $\hat c=2$ and,
hence, is critical in a space of
ultra--hyperbolic signature $(-,-,+,+)$ or in a four--dimensio\-nal Euclidean space.
Possessing intrinsic complex structure it breaks manifest $SO(2,2)$ Lorentz invariance
(for reviews see e.g. Refs. \cite{marc,olaf}). As was demonstrated in Ref. \cite{ov}, at the tree level
the quantum dynamics of $N=2$ string\footnote{Throughout the paper we discuss the closed string case.}
is governed by the Plebanski equation \cite{pleb} and the only
quantum state in the spectrum can be identified with the K\"ahler potential of a Ricci--flat K\"ahler metric
(self--dual gravity).

In fact, any $N=2$ superconformal theory with $\hat c=2$ reveals a
higher $N=4$ symmetry as the spectral flow operator and its inverse can be taken to be the raising and
lowering operators of the $su(1,1)$ subalgebra of a small $N=4$ superconformal algebra (for details see e.g. Ref. \cite{nb}).
Using this observation Berkovits and Vafa demonstrated in Ref. \cite{bv} that the critical $N=2$ string can
be embedded in a more universal $N=4$ topological string framework. As there are no $N=2$ ghosts around \cite{bv},
$N=2$ string scattering amplitudes can be reproduced in a much simpler way. Moreover, the $N=4$ topological
prescription allows one to prove the vanishing theorems to all orders in perturbation theory, which otherwise
are extremely difficult to demonstrate \cite{bv}--\cite{ov_1}. For a related work see Refs. \cite{bvw}, \cite{bv_1}.
Some recent applications are discussed in Ref. \cite{nv}.

Another appealing point concerning the $N=4$ formalism is that it allows one to keep manifest $SO(2,2)$ Lorentz invariance,
at least at the level of classical considerations. As the global automorphism group of a small $N=4$ superconformal algebra
is $SU(1,1) \times SU(1,1)'\simeq SO(2,2)$, the corresponding string theory action functional is manifestly Lorentz
invariant \cite{bg2}. It should be remembered, however, that the classical Lorentz symmetry holds at the price of the functional dependence
of the currents providing a representation of the $N=4$ algebra \cite{ws,bg1}. In the quantum theory the topological prescription of
Ref. \cite{bv} relies upon twisting the $N=2$ algebra by the $U(1)$ current which amounts to choosing a specific complex structure
in the twistor space of all complex structures \footnote{Besides, special care is to be taken of the instanton contribution
to the partition function of the $N=4$ topological theory \cite{bv}.}, thus reducing $SO(2,2)$ to $U(1,1)$ and
reproducing the earlier results of Ref. \cite{ov}. Alternatively, working within the framework of the old covariant
quantization, one discovers that the Lorentz invariance of the boson emission vertex proves to be incompatible with the causality and
cyclic symmetry of tree level scattering amplitudes \cite{bg3}, thus reproducing, once again, the result of Ooguri and Vafa \cite{ov}.

As was mentioned above, the closed $N=2$ string requires a Ricci--flat K\"ahler space as a consistent background.
It is well known that in four dimensions the Ricci--flatness of a K\"ahler manifold implies the
hyper--K\"ahler structure. Exhibiting one complex structure in an explicit form
the conventional formulation of $N=2$ string seems to hide two more complex structures.
It is the purpose of this paper to make the hyper--K\"ahler structure intrinsic to $N=2$ quantum string
in a curved space manifest. Making recourse to the equivalent $N=4$ topological formalism
we demonstrate that, apart from a background metric,
the superconformal currents couple to a background two--form. By superconformal symmetry the latter
is constrained to be holomorphic and covariantly constant and allows one to construct the
missing complex structures in an explicit form, thus demonstrating the advantage of working within the
framework of the $N=4$ formalism.

The organization of the work is as follows. In the next section we examine a small $N=4$ superconformal algebra in
a curved space at the classical level. A generalization of the Poisson bracket which is compatible with the minimal interaction
criterion and Jacobi identities is given. Background fields which allow one to construct a representation of the $N=4$ superconformal algebra
in a curved space include a Hermitian metric and a background two--form. We then show that the algebra closes under
the Poisson bracket, provided the background two--form is holomorphic and covariantly constant. Three complex structures obeying
a (pseudo)quaternion algebra are given in an explicit form. In Sec. III we extend our analysis to the quantum theory and construct a representation
of the small $N=4$ superconformal algebra in a curved space at the tree level. We summarize our results in the concluding Sec. IV.

\noindent

\vskip 0.5cm
{\bf 2. Classical considerations}

\noindent

\vskip 0.5cm
Our classical analysis begins with the simplest representation
of $d=2$, $N=4$ superconformal algebra in a flat space
\bea\label{flatcurrents} &&
T=\textstyle{\frac 12} (p_b \eta^{\bar a b} +\partial_1 x^{\bar
a})(p_{\bar a} +\partial_1 x^b \eta_{b \bar a}) -\textstyle{\frac
{i}{2}}(\p^a\partial_1 \p^{\bar a} +\p^{\bar a}
\partial_1 \p^a) \eta_{a\bar a},
\nonumber\\[2pt]
&& G=(p_{\bar b} \eta^{\bar b a} +\partial_1 x^a)\p^{\bar a}
\eta_{a\bar a}, \quad {\bar G}=(p_b \eta^{\bar a b} +\partial_1
x^{\bar a})\p^a \eta_{a\bar a},
\nonumber\\[2pt]
&& H=(p_{\bar b} \eta^{\bar b a} +\partial_1 x^a)\p^c
\epsilon_{ac}, \quad {\bar H}=(p_b \eta^{\bar a b} +\partial_1
x^{\bar a})\p^{\bar c} \epsilon_{\bar a \bar c},
\nonumber\\[2pt]
&& J=\p^{\bar a} \p^a \eta_{a\bar a}=0,\quad
J^{(1)}=\p^a \p^c \epsilon_{ac}, \quad
J^{(2)}=\p^{\bar a} \p^{\bar c} \epsilon_{\bar a \bar
c}.
\eea
This is constructed on a phase space spanned by a complex
boson $x^a(\tau,\s)$, the conjugate momentum $p_a(\tau,\s)$, and a self-conjugate complex
fermion $\psi^a(\tau,\s)$. We use the ordinary Poisson bracket
\bea
&&
\{x^a(\s),p_c(\s') \}={\d^a}_c \d (\s-\s'), \quad \{x^{\bar
a}(\s),p_{\bar c}(\s') \}={\d^{\bar a}}_{\bar c} \d (\s-\s'),
\nonumber\\[2pt]
&&
\{\p^a(\s),\p^{\bar c}(\s') \}=i \eta^{\bar c a}\d
(\s-\s'),
\eea
and conjugate on the cylinder as ${(x^a)}^{*}=x^{\bar a}$,
${(\p^a)}^{*}=\p^{\bar a}$. In the equation above $\epsilon_{ac}$ is the Levi-Civita
antisym\-metric tensor $\epsilon_{ac}=-\epsilon_{ca}$, $\epsilon_{01}=-1$,
${(\epsilon_{ac})}^{*}=\epsilon_{\bar a \bar c}$ and $\partial_1=\frac{\partial}{\partial\s}$.
We assume periodic boundary conditions for the bosonic fields. For the fermions one can choose
the $NS$ representation due to the spectral flow \cite{ss}.

Cancellation of the conformal anomaly in the quantum theory requires a target space of two complex
dimensions. Depending on the
choice of the original flat metric ${ds}^2=\eta_{a \bar a } dx^a \otimes dx^{\bar a}$,
in real coordinates one reveals either a four-dimensional Euclidean space
or a space of ultra hyperbolic signature. In this paper we stick
to the latter option and henceforward put $\eta_{a\bar a}=\it{diag}(-,+)$,
$a,\bar a=0,1$.

Passing to a curved space\footnote{Our convention is to use $a$, $b$, $c$ indices for the flat case and $k$, $n$, $m$ ones in the presence of a nonvanishing curvature.} (with metric $g^{\bar n n}(x,\bar x)$), first one has to decide which bracket to use. A bracket which
respects Jacobi identities and is compatible with the minimal interaction criterion
can be constructed by going to a larger phase space which involves two canonical pairs
$(\p^n,\pi_n)$, $(\p^{\bar n},\pi_{\bar n})$ and imposing there two second class constraints
\be
\pi_n+\textstyle{\frac{i}{2}} \p^{\bar n}
g_{n\bar n}=0, \quad \pi_{\bar n}+\textstyle{\frac{i}{2}}
 \p^n g_{n\bar n}=0.
\ee
These remove the auxiliary fields $(\pi_n,\pi_{\bar n})$ and leave one with the Dirac bracket
\bea
&&
\{x^N(\s),p_M(\s') \}={\d^N}_M \d (\s-\s'), \quad
\{\p^n(\s),\p^{\bar n}(\s') \}=i g^{\bar n n}\d
(\s-\s')
\nonumber\\[2pt]
&&
\{ p_N (\s), \p^m(\s') \} =\textstyle{\frac {1}{2}}
\p^k \partial_N g_{k \bar s} g^{\bar s m} \d (\s-\s'),
\nonumber\\[2pt]
&&
\{ p_N (\s), \p^{\bar m}(\s') \} =\textstyle{\frac
{1}{2}} \p^{\bar k} \partial_N g_{s \bar k} g^{\bar m s}
\d (\s-\s'),
\nonumber\\[2pt]
&&
\{ p_N (\s), p_M (\s') \}= \textstyle{\frac
{i}{2}} {\p}^k {\p}^{\bar k} g^{\bar s s} \partial_{[N} g_{k \bar s}~ \partial_{M]}
g_{s \bar k}~\d (\s-\s'),
\eea
where $N$ is a collective index for $(n,\bar n)$ and $A_{[N} B_{M]}=\textstyle{\frac 12} (A_N B_M
-A_M B_N)$. Because the $N=4$ algebra in a flat space is essentially complex it
seems natural to preserve the complex structure when switching an external field on. Thus, we take
the background metric to be hermitian $g_{nm}=g_{\bar n \bar m}=0$,
${(g_{n \bar m})}^{*}=g_{m \bar n}$.

Then one has to couple the generators to background fields. Taking into account
commuta\-ti\-on relations characterizing the $N=4$ superconformal algebra (we use the notation in Ref.~\cite{bg3})
it suffices to fix
$G$, $\bar G$ and the $R$--symmetry generators $J$, $J^{(1)}$, $J^{(2)}$.

Assuming the coupling to be minimal
($\partial_\a \p^n~\rightarrow~\nabla_\a \p^n = \partial_\a \p^n +\partial_\a x^p {\G^n}_{ps}
\p^s$)
\be
p_n \rightarrow p_n
+\textstyle{\frac i2}\p^{\bar m} \p^s {\G^k}_{ns}
g_{k\bar m} \equiv \Pi_n, \quad
p_{\bar n} \rightarrow p_{\bar n}
-\textstyle{\frac i2} \p^{\bar m} \p^s {\G^{\bar
k}}_{\bar n \bar m} g_{s\bar k} \equiv \Pi_{\bar n},
\ee
the first three currents are easily constructed
\be\label{currents1}
G=(\Pi_{\bar n} +\partial_1 x^n g_{n\bar
n})\p^{\bar n}=0, \quad {\bar G}=(\Pi_n +\partial_1 x^{\bar
n} g_{n\bar n})\p^n=0,
\quad J=\p^{\bar n} \p^n g_{n\bar n}=0.
\ee

As the metric carries one holomorphic index and one antiholomorphic index, the change $\epsilon_{nm}
\rightarrow \epsilon_{nm}/\sqrt{-\det g}$ commonly accepted in real spaces does not yield a tensor field.
Thus, in order to formulate $J^{(1)}$ and $J^{(2)}$ one is forced to introduce into consideration a background two--form
$B_{nm}={(B_{\bar n \bar m})^{*}}$ which reduces to $\e_{nm}$ in a flat limit.
With this at hand one can set
\be\label{currents2}
J^{(1)}=\p^n \p^m B_{nm}, \quad J^{(2)}=\p^{\bar
n} \p^{\bar m} B_{\bar n \bar m}.
\ee

Now it is important to notice that the nilpotency of the
supersymmetry charge $G$ holds only if the background metric
is k\"ahlerian
\be\label{constr}
\partial_n g_{m \bar m}- \partial_m g_{n \bar m}=0, \qquad
\partial_{\bar n} g_{m \bar m}- \partial_{\bar m} g_{m \bar n}=0.
\ee
This means, in particular, that the connections ${\G^k}_{n p}$ and ${\G^{\bar k}}_{\bar n \bar p}$
become symmetric and $(\Pi_n,\Pi_{\bar n})$ in Eq. (\ref{currents1}) above can be reduced to
$(p_n,p_{\bar n})$.

Finally, it is a matter of a straightforward although a bit lengthy calculation to verify that the entire
$N=4$ superconformal algebra closes provided the remaining currents have the form
\bea
&&
H=(\Pi_{\bar k} g^{\bar k n} +\partial_1
x^n) \p^m B_{nm}, \quad {\bar H}=(\Pi_k g^{\bar n k}
+\partial_1 x^{\bar n}) \p^{\bar m} B_{\bar n \bar m},
\nonumber\\[2pt]
&&
T=\textstyle{\frac 12} (\Pi_{\bar n}+\partial_1 x^p g_{p \bar
n})(\Pi_n +\partial_1 x^{\bar p} g_{n \bar p} ) g^{\bar n n}
-\textstyle{\frac {i}{2}}(\p^n \partial_1 \p^{\bar n}
+\p^{\bar n} \partial_1 \p^n) g_{n\bar n}-
\nonumber\\[2pt]
&&
\qquad -\textstyle{\frac {i}{2}} \p^{\bar n} \p^n
\partial_1 x^p {\G^k}_{n p} g_{k \bar n}
+\textstyle{\frac {i}{2}} \p^{\bar n} \p^n
\partial_1 x^{\bar p}
{\G^{\bar k}}_{\bar n \bar p} g_{n \bar k}
\eea
and the background two--form obeys the
restrictions
\be\label{hypkael1}
\partial_{\bar
k} B_{nm}=0, \quad \nabla_k B_{nm}=0.
\nonumber\\[2pt]
\ee
In checking the algebra the integrability conditions
\bea
&&
{R^k}_{n \bar m s} B_{kp} -{R^k}_{n \bar m p} B_{ks} =0, \quad
{R^{\bar k}}_{\bar n m \bar s} B_{\bar k \bar p} -{R^{\bar
k}}_{\bar n m \bar p} B_{\bar k \bar s} =0,
\eea prove to be helpful. Besides, one has to use
the algebraic relation $B_{\bar n \bar m} B_{sp} g^{\bar m s}=g_{p \bar n}$ which holds true
for an irreducible manifold~\cite{yano}.

Thus, in order to support a small $N=4$ superconformal algebra a background K\"ahler manifold
must admit a covariantly constant holomorphic two--form. As is well
known~\cite{lichn}, this reduces the holonomy group of a manifold
to a subgroup of $SU(1,1)$ which implies a pseudo--hyper--K\"ahler space.
Indeed, along with a natural complex structure characterizing the case
\be J=\left(\begin{array}{cc}
i{\d_m}^n & 0\\
0 & -i{\d_{\bar m}}^{\bar n} \\
\end{array}\right), \quad J^2=-1,
\ee
one can construct two real structures
\be
S=\left(\begin{array}{cc}
0& B_{nm}g^{\bar k m}\\
B_{\bar n \bar m} g^{\bar m k} & 0 \\
\end{array}\right), \quad
T=\left(\begin{array}{cc}
0& iB_{nm}g^{\bar k m}\\
-iB_{\bar n \bar m} g^{\bar m k} & 0 \\
\end{array}\right), \quad S^2=1, \quad T^2 =1,
\nonumber\\[2pt]
\ee
altogether forming a pseudo--quaternionic algebra (in this respect see also Refs. \cite{gp}, \cite{ch})
\be
ST=-TS=-J, \quad TJ=-JT=S,
\quad JS=-SJ=T.
\ee

If the original flat metric were chosen in the form $\eta_{n\bar n}=\mbox{diag}(+,+)$,
an ordinary quaternionic algebra and hyper--K\"ahler geometry would be reproduced in this way.
Worth mentioning also is that a (pseudo) hyper--K\"ahler space is automatically Ricci flat.
Indeed, it suffices to contract the integrability condition
${R^k}_{n \bar m s} B_{kp} -{R^k}_{n \bar m p} B_{ks} =0$ with the tensor
$g^{\bar l p} B_{\bar l \bar r} g^{\bar r s}$, in order to get $R_{n{\bar n}}=0$.

\noindent

\vskip 0.5cm
{\bf 3. Tree-level quantum consideration}

\noindent

\vskip 0.5cm

We next wonder if the restrictions on background geometry derived
in the previous section allow one to construct a quantum representation of a small $N=4$ superconformal
algebra. By now only a perturbative technique is
available, this appealing to the use of Riemann coordinates (see e.g. \cite{afm}).
As the geometry characterizing the case is essentially complex, the passage to Riemann coordinates
should be realized by a holomorphic transformation \cite{gaume2}.

Thus, instead of using a
geodesic connecting an arbitrary point and the origin one performs the following classical--quantum splitting
\bea
&&
x^n=x_0^n+\xi^n -\frac{1}{2!} {\Gamma^n}_{ml} (x_0,{\bar x}_0) \xi^m \xi^l -\frac{1}{3!} {\hat\nabla}_p
{\Gamma^n}_{ml} (x_0,{\bar x}_0) \xi^p \xi^m \xi^l +\dots,
\nonumber\\[2pt]
&&
\psi^n=\frac{\partial x^n}{\partial \xi^m} \lambda^m=\lambda^n -{\Gamma^n}_{ml} (x_0,{\bar x}_0) \xi^m \lambda^l+\dots~.
\eea
This renders a background field expansion covariant with respect to holomorphic changes of coordinates.
The covariant derivative ${\hat\nabla}_p$ entering the last line acts only on lower indices and the
fermionic field is taken to be purely quantum. It is assumed that the point
lies in the normal neighborhood of the origin, i.e. the exponential map from the tangent space to the origin
onto the neighborhood is the diffeomorphism. So, two geodesics passing through the origin do not intersect in other
points of the neighborhood.

In order to define quantum propagators, one then introduces a complex zweibein ${e_n}^a$ and represents the metric
and the background two--form as follows:
\be\label{zw}
g_{n\bar n}={e_n}^a {e_{\bar n}}^{\bar a} \eta_{a\bar a}, \qquad B_{nm}={e_n}^a {e_m}^b \e_{ab}.
\ee
It is assumed that the zweibein is covariantly constant
and that the spin connection one--forms $dx^n {{\omega_n}^a}_b$, $dx^{\bar n} {{\omega_{\bar n}}^a}_b$,
$dx^n {{\omega_n}^{\bar a}}_{\bar b}$, $dx^{\bar n} {{\omega_{\bar n}}^{\bar a}}_{\bar b}$ take
values in the Lie algebra $su(1,1)$
\bea
&&
{{\omega_N}^a}_b \eta_{a \bar b} + {{\omega_N}^{\bar a}}_{\bar b} \eta_{b \bar a}=0, \quad
\quad
{{\omega_N}^a}_a={{\omega_N}^{\bar a}}_{\bar a}=0,
\nonumber\\[2pt]
&&
{{\omega_N}^c}_a \e_{cb}-{{\omega_N}^c}_b \e_{ca}=0, \quad {{\omega_N}^{\bar c}}_{\bar a} \e_{\bar c \bar b}
-{{\omega_N}^{\bar c}}_{\bar b} \e_{\bar c \bar a}=0.
\eea

As the background metric is hermitian it contains four real components.
The balance between these four degrees of freedom and eight real components of
the zweibein is provided by the local $U(1,1)$ transformation ${{e'}_n}^a={\Lambda^a}_b {e_n}^b$
with four real parameters which leaves the metric invariant. Specifying a
covariantly constant two--form like in Eq. (\ref{zw}) above, one breaks the local
$U(1,1)$ symmetry down to $SU(1,1)$ and destroys the equilibrium. This can be reestablished
by imposing one real equation
\be
\det g_{n\bar n}=-\Omega(x) \bar\Omega(\bar x),
\ee
where $\Omega(x)$ is a fixed holomorphic function and $\bar\Omega(\bar x)$ is its complex conjugate.
This is fully consistent with the Ricci--flatness of a background manifold
because $R_{n\bar n}=\partial_n \partial_{\bar n} \ln (\det g_{m \bar m})$. In particular, taking
$\Omega=1$ one recovers the Plebanski equation \cite{pleb} which implies also that locally
the background two--form is a constant
(for a related discussion see also Ref. \cite{ov}).

Redefining the quantum fields $\xi^a=\xi^n {e_n}^a$, $\lambda^a=\lambda^n {e_n}^a$, one can further specify the
propagators (we Wick rotate the temporal coordinate on the world--sheet and go over to a complex plane)
\be
\langle \xi^a (z,\bar z) \xi^{\bar a} (w,\bar w)\rangle=-\eta^{\bar a a} (\ln(z-w)+\ln (\bar z -\bar w)),
\qquad
\langle \lambda^a (z) \lambda^{\bar a} (w) \rangle=-\frac{\eta^{\bar a a}}{z-w}.
\ee
With these at hand one can decompose the conformal currents in Riemann coordinates and evaluate operator product
expansions perturbatively. Unfortunately, decomposing generators in this way one does not arrive at
a closed algebra and extra terms have to be added to the currents in order to close the algebra \cite{bns}.

At the tree level a necessary modification is prompted by the algebra itself. In par\-ti\-cular,
taking the linear approximation for $G$, $\bar G$ and the $R$--symmetry generators
(here we denote $\partial x_0^a=\partial x_0^n {e_n}^a$)
\bea
&&
G=(\partial x_0^a+\nabla \xi^a ) \eta_{a\bar a} \lambda^{\bar a}+\dots, \quad
\bar G=(\partial x_0^{\bar a}+\nabla \xi^{\bar a}) \eta_{a\bar a} \lambda^a+\dots,
\nonumber\\[2pt]
&&
J=\lambda^{\bar a} \lambda^a \eta_{a\bar a}+\dots, \quad J^{(1)}=\lambda^a \lambda^b \e_{ab}+\dots, \quad
J^{(2)}=\lambda^{\bar a} \lambda^{\bar b} \e_{\bar a \bar b}+\dots,
\eea
where $\nabla\xi^a=\partial\xi^a+\partial x_0^N {{\omega_N}^a}_b \xi^b$, one can fix the remaining generators
\bea
&&
T=-(\partial x_0^a+\nabla \xi^a ) (\partial x_0^{\bar a}+\nabla \xi^{\bar a}) \eta_{a\bar a}+
\frac{1}{2}(\lambda^{\bar a} \nabla \lambda^a +\lambda^a \nabla \lambda^{\bar a} )\eta_{a\bar a}
+\lambda^{\bar a} \lambda^a \partial x_0^N {{\omega_N}^b}_a \eta_{b \bar a}+\dots,
\nonumber\\[2pt]
&&
H=(\partial x_0^a+\nabla \xi^a )\lambda^b \e_{ab}+\dots, \quad \bar H=(\partial x_0^{\bar a}+\nabla \xi^{\bar a})
\lambda^{\bar b} \e_{\bar a \bar b}+\dots,
\eea
and verify after a tedious calculation that the entire algebra closes. When checking the algebra divergent terms appear which
are handled by dimensional regularization \cite{bns}.

Thus, at the tree level the only correction to the naive decomposition of currents
is given by the term $\lambda^{\bar a} \lambda^a \partial x_0^N {{\omega_N}^b}_a \eta_{b \bar a}$ which
enters the conformal generator $T$. Curiously enough, this breaks manifest $U(1,1)$ local
invariance (for a related discussion see Ref. \cite{cfmp} and Refs. \cite{sb3}--\cite{sb2}).

\noindent

\vskip 0.5cm
{\bf 4. Conclusion}

\noindent

\vskip 0.5cm

To summarize, in the present paper we considered a small $N=4$ superconformal algebra in a curved space.
Our analysis reveals an interesting interplay between the maximally extended superconformal algebra admitting unitary representations
and hyper--K\"ahler geometry. In particular, a covariantly constant holomorphic two--form which specifies the holonomy group
of a target manifold explicitly couples to the superconformal currents.
On the string theory side, our analysis suggests that the $N=4$ topological formalism yields the appropriate framework
to keep the hyper--K\"ahler structure intrinsic to the $N=2$ quantum string manifest.

Although we did not try to systematically extend the present consideration to the one--loop order and beyond,
according to the analysis of Refs. \cite{ov}, \cite{bv}--\cite{ov_1} no quantum corrections to the equations specifying the
background fields will follow (see, however, Ref. \cite{bgi}). It would be interesting to check this by explicit calculations.
Preliminary considerations show, however, that a naive decomposition of $G$ in Riemann coordinates fails to yield
$G \cdot G \sim 0$, so $G$ itself should be modified appropriately.

\vspace{0.5cm}
\noindent
{\bf Acknowledgment}\\[-4pt]

\noindent
We thank Joseph Buchbinder for useful discussions. This work was partially supported by NATO
Collaborative Linkage Grant PST.CLG. 97938,
INTAS-00-00254 grant,
RF Presidential grants MD-252.2003.02, NS-1252.2003.2, INTAS grant 03-51-6346,
RFBR-DFG grant 436 RYS 113/669/0-2, RFBR grant 03-02-16193 and the European Community's Human Potential
Programme under contract HPRN-CT-2000-00131 Quantum Spacetime.

\end{document}